**Pre-Print.**

## ABSTRACT

The United Arab Emirates (UAE) is a young, oil-rich country, where national youth display a clear preference for public sector employment. Growing youth unemployment reinforces the importance of non-government employment, including entrepreneurship. This study investigates UAE national youth intentions towards entrepreneurship through the Theory of Planned Behavior and the Entrepreneurial Intention Questionnaire (EIQ). Analysis (N=544) identifies the direct influence of attitude and perceived behavioral control, and indirect influence of subjective norms on entrepreneurship intention. Results also examine several demographic variables and highlight the potential importance of family and social groups in promoting entrepreneurial intentions in this emerging country context.

## Introduction

The central purpose of the current study is to gain a better understanding of some of the many factors that may influence the choice of entrepreneurship as a career option among a population of UAE national youth. The current paper is partly an extraction and theoretical refinement of work originally conducted for the award of Doctorate of Business Administration (Al Saiqal 2017). Entrepreneurship is important for achieving economic development, growth, diversification and internationalization, as well as job creation (Gallant, Majumdar, & Varadarajan, 2010; Mihailova, Shirokova, and Laine 2015; Van Gelderen et al., 2008). Although its definition varies among researchers, this study adopts the general definition of entrepreneurship as "starting one's own business" and thus becoming self-employed (Engle et al., 2010; Van Gelderen et al., 2008). While much business and entrepreneurship research takes place in western contexts, there is a growing call for the examination of business phenomena in non-western contexts (Gaur, Kumar & Singh, 2014; Singh & Gaur, 2013). The specific challenge of entrepreneurship in an emerging country context is of particular interest given an often ambiguous and uncertain business environment (Singh, & Gaur, 2018).

The issue of entrepreneurship in the UAE is particularly important in an employment environment dominated by discussion of work force nationalization in the public and private sectors. In such a context entrepreneurship is considered as an underutilized "*third stream*" career option for UAE national youth (Ryan, Tipu, & Zeffane, 2011, p. 154). Recent estimates place UAE national youth unemployment at nearly 25 percent, which is considered one of the highest in the world (Horne, Huang, & Awad, 2011; Sokari, Horne, Huang, & Awad, 2013). In many emerging economy contexts governments exert strong market influences (Gaur, Ma, & Ding, 2018; Singh, Pattnaik, Gaur, & Ketencioglu, 2018). In the UAE such influences extend heavily into business and labor markets, affecting entrepreneurial activity. It is also suggested that transitioning from an of oil-based economy offers additional unique challenges to economic development and entrepreneurship activity (Kaynak, & Nasirova, 2005).

UAE national youth are known to prefer working in the already saturated government sector, avoiding the high-pressure working environments of the private sector (Ryan 2016; Salih, 2010). Moreover, it is suggested that private sector employers have a preference for the lower expectations of expatriate workers with higher skills. As a result, many government and workforce nationalization (Emiratization) initiatives have achieved only modest results in addressing unemployment levels among UAE nationals (Harry, 2007; Ryan et al., 2011; Ryan, 2016; Salih, 2010). It is argued that, to date, Emiratization policies focus largely on improving national skills, education and qualifications while neglecting the need to influence and reshape attitudes, beliefs and opinions regarding work in the private sector, and/or self-employment (Daleure, Albon, & Hinkston, 2014).

The Global Entrepreneurship Monitor (GEM, 2011) reports that although a very high proportion (51.9%) of UAE youth perceive opportunities for entrepreneurial activity, few take the necessary steps to seize these opportunities. Despite a strong work ethic supportive of entrepreneurial activity, few UAE youths pursue entrepreneurship as a career choice (Ryan, & Tipu, 2016). Research also suggest that the link between an entrepreneurial orientation and actual performance is significantly moderated by the entrepreneurship context and environment (Bogatyreva et al., 2017). This highlights the importance of examining entrepreneurial phenomena in different contexts such as the UAE. Currently, a very low proportion of UAE youth are involved in early-stage entrepreneurial activities and intention to start new businesses in the UAE is low, at only 2% (Horne, Huang, & Awad, 2011). The reasons for this low rate are suggested as (a) the economic cost of failure, which indicates the loss that would be incurred by business failure in terms of monetary, financial and other tangible resources; (b) the social cost of failure, which is related to loss of reputation, shame to one's family and embarrassment; and (c) the personal cost of failure, which indicates how individual business failure affects motivation levels, perceived personal abilities, capabilities, skills and intelligence. Furthermore, a fear of failure may be reinforced by inadequacies in the UAE legislative framework and entrepreneurship ecosystem.

The UAE government has launched initiatives to prepare the country for the post-oil era, such as "Absher" in 2013, with goals of enhancing national participation in the workforce and expanding career options for UAE nationals. Moreover, the UAE national agenda 2021 stresses the need to develop and enhance entrepreneurship attitudes, activities and aspirations. In spite of the importance of entrepreneurship in supporting government aspirations for its citizens, research on youth intention or "would be" entrepreneurs in the UAE context is severely lacking (Ryan, Tipu, & Zaffane, 2011). Accordingly, the present study focuses on entrepreneurial intention in the UAE. A systematic review of the literature, as advised by Gaur and Kumar (2018), identifies the relevant theoretical and conceptual dimensions for inclusion in the current study. A more thorough understanding of the issue is thus gained by exploring the factors that may influence the choice of entrepreneurship as a career option among the population of UAE national university students who represent potential future entrepreneurs (Gallant et al., 2010; Matlay et al., 2012; Turker & Selcuk, 2009).

**Literature Review and Hypothesis Development**

Entrepreneurial intention is defined as the "intention of setting up one's business in the future" and it involves a process of prior planning and thinking (Schlaegel & Koenig, 2014; Van Gelderen et al., 2008 p. 5). Much recent study of entrepreneurial behaviors view entrepreneurship as a combination of mental processes (thoughts) and interactions with the external environment (García, González, & García, 2011; Krueger, 2003; Sikdar & Vel, 2011).

Investigating and understanding entrepreneurial intentions improves our understanding of the way in which entrepreneurs are developed and emerge (Engle et al., 2010; Solesvik, 2017).

Intention is considered to be a significant, valid, and unbiased predictor of career choice, where intention-based models provide a deep and practical insight on planned behavior (Almobaireek & Manolova, 2012; Krueger, Reilly, & Carsrud, 2000). The Theory of Planned Behavior (TPB), one of the most widely used intention models and provides a framework for studying and understanding entrepreneurial intention, is based on three constructs, namely Attitude (referring to the degree to which individuals perceive the attractiveness of the behavior in question), Subjective Norm (SN) (referring to the perceived social pressure to perform the behavior from significant others; such as family, friends, role models, and others) and Perceived Behavioral Control (PBC) (referring to the self-evaluation of one's own competence with regard to the task or behavior) (Ajzen, 1991). These three main specifications represent individuals' experiences and observations, which in turn act as a foundation on which to develop three different "salient" beliefs –behavioral beliefs, normative beliefs and beliefs drawn from experience (Engle et al., 2010). The TPB model provides a comprehensive framework to analyze entrepreneurial intention and is well suited to explain entrepreneurship which is opportunity driven as opposed to necessity driven -as is mostly the case with entrepreneurship among UAE nationals (Sokari et al., 2013). Moreover, researchers suggest that the TPB could be used as a *"culture-universal theory",* to predict career intentions in any country (Moriano et al., 2011, p. 2). It is argued that the more favorable the attitude and subjective norm, and the greater the perceived behavioral control, the stronger is the intention to perform the directed behavior (Ajzen, 1991; Autio et al., 2001; Matlay et al., 2012).

Many researchers support the direct and positive effects of attitude and perceived behavioral control on entrepreneurial intentions in different cultures and contexts (Autio et al., 2001; Liñán & Chen, 2009; Matlay et al., 2012; Moriano et al., 2011; Rueda et al., 2015; Van Gelderen et al., 2008). Consequently, we propose the following hypotheses.

***H1: Attitude towards entrepreneurship positively and directly influences entrepreneurial intention.***

***H2: Perceived behavioural control positively and directly influences entrepreneurial intention.***

However, subjective norm effects are rarely supported and appear to have insignificant and weak direct relationship with entrepreneurial intention (Armitage & Conner, 2001; Kautonen et al., 2015; Liñán & Chen, 2009; Matlay et al., 2012; Moriano et al., 2011; Schlaegel & Koenig, 2014).

Liñán & Chen (2009) suggested that in order to fully understand the effect of subjective norm on entrepreneurship intention, there is a need to investigate the indirect effect of this construct. That is, there is a need to investigate its direct effects on both attitude and perceived behavioral control. The UAE can be described as a hierarchical culture (Al-Ali, Singh, Al-Nahyan, & Sohal, 2017). In such a context we might expect the influence of family and peer groups to be significant. Indeed, given the unique context of the UAE and based on available literature, it may be the case that subjective norm has a powerful direct and indirect effect on entrepreneurial intention in the UAE context. It is considered that this antecedent may be highly influenced by the role of families in UAE culture as they play an important role in shaping youth career preferences and choice (Khalifa Fund, 2014; Carr & Sequeira's, 2007). Based on this previously reviewed literature relating to the concept of subjective norm the current study hypothesizes the following.

***H3: Subjective norm positively and influences entrepreneurial intention.***

***H4: Subjective norm positively influences attitude towards entrepreneurship.***

***H5: Subjective norm positively influences perceived behavioural control.***

Although attitude, subjective norm and perceived behavioral control represent the theoretical constructs influencing intention, researchers have also argued that other variables might indirectly affect the entrepreneurship intention through direct effects on the theory of planned behaviors three main constructs. These variables can include entrepreneurship education, age, gender, self-employment experience and entrepreneurial role models (Almobaireek & Manolova, 2012; Kolvereid & Isaksen, 2006; Liñán & Chen, 2009; Tipu, Zeffane, & Ryan, 2011; Yang, 2013).

Education in general represents an important means of developing human capital for any economic endeavors. Universities can foster entrepreneurship activity by enhancing the EI among students through directing their career preferences and increasing entrepreneurial awareness. Entrepreneurship education is positively related to entrepreneurial intention especially in collectivist cultures (Bae et al., 2014). Consequently, the provision of quality entrepreneurship training can raise the tendency among youth to become entrepreneurs (Almobaireek & Manolova, 2012; Liñán & Chen, 2009; Matlay et al., 2012). In their meta-analysis of entrepreneurship education outcomes, Martin, McNally, & Kay (2013) confirm that entrepreneurship education is associated with higher levels of intention to become an entrepreneur. Moreover, based on the survey conducted on fifty entrepreneurs from diverse backgrounds in selected universities across the UAE, Kargwell & Inguva (2012) find many entrepreneurs believe that education is a critical success factor in their business.

Education within the UAE context might be assumed to have a direct effect on young people's attitude and PBC, but perhaps not on subjective norm. While family involvement has been shown to influence academic achievement (Alhosani, Singh, & Al Nahyan, 2017), families in the UAE are often not fully involved in the educational programs and other awareness initiatives which aim to promote entrepreneurship among UAE youth. However, families' opinions regarding career choice (which largely encourages public sector employment) are highly respected by UAE national youth. Accordingly, this factor may affect the entrepreneurial intention indirectly through its direct influence on attitude and perceived behavioral control. It is argued that little can be done by educational institutions or policy makers to influence UAE national youth career choices if their families actively encourage them toward working only in the public sector (Daleure, Albon, & Hinkston, 2014; Khalifa Fund, 2014). Consequently, we hypothesize the following:

*H6a: Education directly influences entrepreneurial intention.*

*H6b: Education directly influences attitude towards entrepreneurship.*

*H6c: Education directly influences perceived behavioural control.*

*H6d: Education does not directly influence subjective norm.*

In the UAE context, investigating the relationship between gender and EI is especially important. Gender difference in entrepreneurial activity in emerging country contexts are well documented, (Chatterjee, Dutta Gupta & Upadhyay, 2018), and gender differences in factors affecting entrepreneurial intentions in the UAE have been identified (Tipu and Ryan 2016). There is a general belief that male Emiratis are more likely than females to be involved in entrepreneurial activities due to additional social, cultural and environmental factors that may stand in the way of women who wish to pursue these activities, and thus their choice of entrepreneurship as a career (Kargwell, 2012; Kargwell & Inguva, 2012; Khalifa Fund, 2014). Research suggests that the percentage of Emirati females who expect to start a business is only 6.5%, while for Emirati males it is 12.1% (Sokari, Horne, Huang, & Awad, 2013). Consequently, we hypothesize the following:

*H7a: Gender directly influences entrepreneurial intention.*

*H7b: Gender directly influences attitude towards entrepreneurship.*

*H7c: Gender directly influences, subjective norm.*

*H7d: Gender directly influences perceived behavioural control.*

Several studies also suggest that age and entrepreneurship experience effect entrepreneurial intention and have a direct influence on the attitude, perceived behavioral control and subjective norms (Almobaireek & Manolova, 2012; Hatak, Harms, & Fink 2015; Kolvereid & Isaksen, 2006; Liñán & Chen, 2009; Yang, 2013). Consequently, the current study incudes these variable to fully examine the important dimensions that might affect UAE youth EI. Accordingly, we hypothesize the following:

*H8a: Age directly influences entrepreneurial intention.*

*H8b: Age directly influences attitude towards entrepreneurship.*

*H8c: Age directly influences subjective norm.*

*H8d: Age directly influences perceived behavioural control.*

*H9a: Entrepreneurship experience directly influences entrepreneurial intention*

*H9b: Entrepreneurship experience directly influences attitude towards entrepreneurship.*

*H9c: Entrepreneurship experience directly influences subjective norm.*

*H9d: Entrepreneurship experience directly influences perceived behavioural control.*

Figure (1) below represent the proposed model for the present study. The model is based largely on the traditional model of theory of planned behavior, with additional antecedent variables included, as suggested by relevant literature. Arrows linking variable indicate positive hypothesized directional relationships.

**Figure (1): The Study Model**

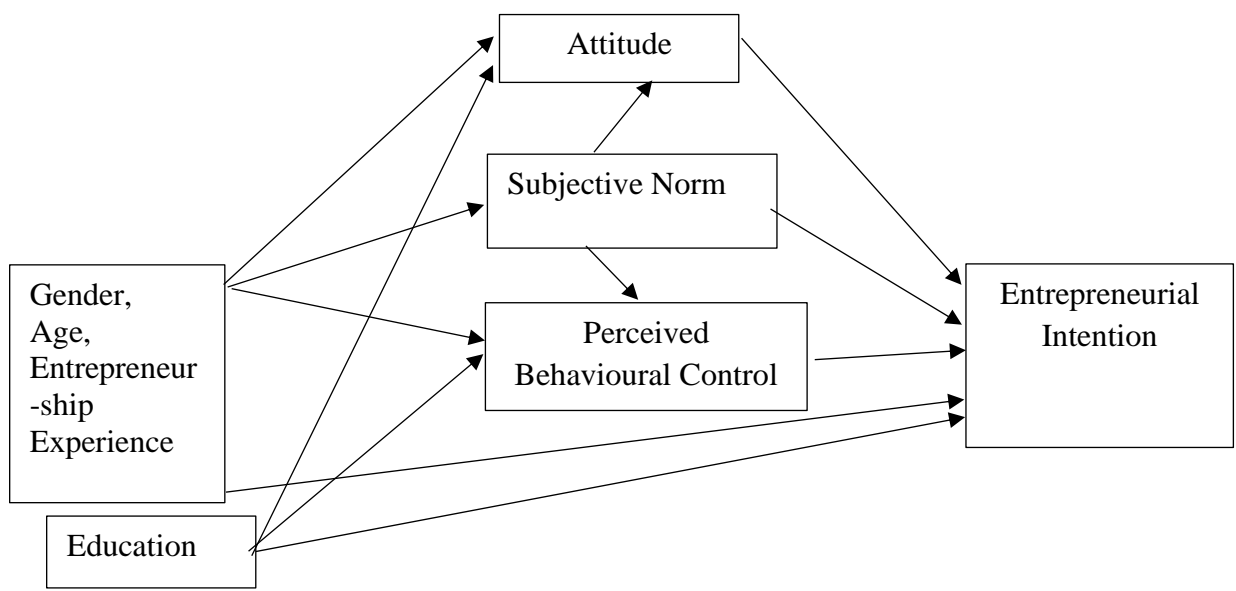

**Methodology and Data**

*Participants*

The present study focuses on the entrepreneurship intentions of UAE national university students. Data collection targeted senior students as they are more knowledgeable of the job market and more engaged in considering employment opportunities, including entrepreneurship (Ahmed et al., 2010; Pratheeba, 2014). Attention was also focused on business and engineering students, as these two areas of study are known to foster many future entrepreneurs (Pratheeba, 2014; Souitaris et al., 2007).

Sampling of students inside the UAE was directed at those institutions which educate the majority of UAE nationals, namely Zayed University (ZU), United Arab Emirates University (UAEU), the Higher Colleges of Technology (HCT) and the Petroleum Institute (PI). Together these key government supported education institutions account for 57% of all enrollments by UAE nationals (Khalifa Fund, 2014).

There are many challenges to conducting research in transitional and emerging country contexts (Michailova, & Liuhto, 2001). In particular, attaining organizational access to participants in an emerging country context can pose unique challenges (Ryan and Daly, 2018). To assist in overcoming organizational access issues, research confederates in each of the aforementioned institutions of the UAE distributed the online survey instrument to the targeted sample via email. Harzing et al. (2005) refer to the significant importance of such confederates in gaining access to research participants in an international business and management context that often presents considerable challenges above and beyond the norms of the research process in a Western context. Surveying of UAE national students studying overseas was facilitated by the UAE government who distributed the questionnaire on behalf of the researchers. A total of 544 usable responses were received. Due to the use of confederates to support data collection an exact response rate for administered surveys is unavailable, though in one institution where a complete record of targeted students was made available the response rate was 34%.

The study found that most of the respondents (72.1%) were between 21 and 23 years old. Nearly two-thirds (70.2%) of the respondents were female and (29.8%) were male. This is reflective of the higher proportion of females than males in higher education in the UAE (Madsen & Cook, 2010). Moreover, most of the participants resided in Abu Dhabi (79.2%) with the remainder either in Umm Al Quwain (0.9 %) or in Ajman (1.5%). In addition, most of the respondents (91.9%), equaling 500 respondents in this survey, were taking a degree in the UAE, whereas only 44 (8.1%) respondents were doing so beyond its borders. The USA accounted for most of the latter group (28 respondents, 5.1%). The analysis also shows that 63.2% of the respondents were taking courses with some entrepreneurship component. The majority of participating students (88.2%) had no previous entrepreneurship experience, which was expected in a student population of this type (Zhang et al., 2015).

*Measures*

The current study utilizes the most recent version of the previously validated standardized measure of the Entrepreneurial Intention Questionnaire (EIQ), (Liñán and Chen, 2009; Jaén & Liñán, 2013). The EIQ was developed following Ajzen's methodological recommendations for building a TPB questionnaire (Jaén & Liñán, 2013). It is a flexible measure employing likert-type responses to well formulated statements and has been revised and used successfully by many authors (Liñán, Urbano, & Guerrero, 2011). The EIQ includes a five-item measure of entrepreneurship intention (EI), the dependent variable in the current study. Attitude was measured with two sets of nine items. The items included the six items proposed by Jaén & Liñán (2013). In addition, three items were added to account for unique elements of the UAE context. These items concerned contributing to society and country, pursuing one's passion and creating a job for oneself. The two sets assess first the attitudes towards the expected outcomes of an entrepreneurial career and the second set assesses attitudes towards the desirability of those outcomes. The two sets were then suggested to be multiplied and divided by the total of items used to obtain scale average scores –nine items- (Jaén & Liñán, 2013). This approach is also supported by Moriano et al. (2011).

Subjective norm (SN) was measured with two sets of five items. The items included the three items proposed by Jaén & Liñán (2013). In addition, based on the literature review and considering the UAE context, the authors added one more item to measure the affect of husband/wife, and split the effect of parents and siblings to measure their effects accurately and individually. These two sets are also suggested to be multiplied together and then divided by the number of items to obtain scale average scores (Jaén & Liñán, 2013).

Perceived Behavioral Control (PBC) was also measured through the approach of Jaén & Liñán (2013). This measurement approach combines elements of self-efficacy and controllability "*the extent to which successfully performing the behavior is up to the person*" (Liñán & Chen, 2009, p. 602). Respondents were asked to indicate the extent that they would be able to effectively perform specific business and entrepreneurship related tasks. The original items suggested by Jaén & Liñán (2013) numbered six; however, an additional item was added related to writing business plans, as this item is suggested to have a positive effect on entrepreneurial intentions in the UAE context (Khalifa Fund, 2014). Accordingly, the respondents were asked to indicate on a five-point Likert scale how effectively they could perform tasks such as defining a business idea, controlling the new-venture creation process, negotiating and maintaining favorable relationships with potential investors, writing a business plan, recognizing opportunities in the market, raising capital and launching a new venture.

For the purpose of capturing data relating to possible antecedent variables, additional questions relating to participant gender, age, entrepreneurship education experience, entrepreneurial experience, were also asked. All measurement items were entered into an online survey for ease of distribution to the target population

*Preliminary analysis and data screening*

An initial set of 719 surveys were returned, however 164 cases were removed as they contained many incomplete scale responses. Additionally, a test of Mahalanobis distance identified several cases of extreme outliers, which were also removed, leaving the final 544 participant responses of the current study.

Common method bias was also examined through the use of a Harman's single factor test (Podsakoff, MacKenzie, Lee, & Podsakoff, 2003). As single factor was accounted for more than 50% of variance, common method bias is not viewed as a significant concern (Eichhorn, 2014). As presented in Table 1, initial examinations of scale reliability also identified acceptable Cronbach alpha reliability coefficients ranging from 0.74 to 0.91 (Nunnally, 1978).

Construct reliability above the accepted minimum of 0.70 was also identified, as was an average variance extracted (AVE) greater than 0.5. (Čater & Čater, 2010; Fornell & Larcker, 1981; Hair, et al., 2005; Hooper, Coughlan, & Mullen, 2008; Liang & Wen-Hung, 2004).

**TABLE 1. Reliability Analysis for the Research Constructs**

|  | AVE | Composite Reliability | Cronbach's Alpha |
|---|---|---|---|
| **Attit** | 0.520 | 0.896 | 0.867 |
| **EI** | 0.632 | 0.889 | 0.837 |
| **PBC** | 0.647 | 0.928 | 0.909 |
| **SN** | 0.569 | 0.840 | 0.744 |

Table 2. provides information on the mean and standard deviations for the main constructs as well as a summary of bivariate correlations across these constructs.

**TABLE 2. Mean, Standard deviations and bivariate correlations**

| Mean | SD |  | EI | Attit | PBC | SN |
|---|---|---|---|---|---|---|
| 3.9 | 0.79 | **EI** | 1.00 |  |  |  |
| 4.00 | 0.73 | **Attit** | 0.457 | 1.00 |  |  |
| 4.01 | 0.76 | **PBC** | 0.315 | 0.410 | 1.00 |  |
| 3.7 | 0.82 | **SN** | 0.211 | 0.328 | 0.241 | 1.000 |

## Analysis and Results

The partial least-square approach (PLS) to Structural Equation Modelling (SEM) is used to assess the structural model. PLS is an increasingly popular analytical technique in empirical management research due to its ability to confirm proposed theory and test the existence of relationship within data (Chin, 1998; Chin and Newsted, 1999; Henseler & Chin, 2010). The path modelling and analysis for the current study were performed using SmartPLS software (http://www.smartpls.com). SEM coefficient significance is tested by approximating the standard error using bootstrapping techniques. Bootstrapping is commonly used when the assumptions of ordinary (parametric) statistical tests, such as data normality are not fully met (Hair, Hult, Ringle, & Sarstedt, 2016).

Arrows are used to symbolize the hypothesized relationships and the direction of influence in the model. Figure 2 depicts the proposed structural model, reflecting the relationships between the variables. The value of the path coefficient associated with each path represents the strength of each linear influence.

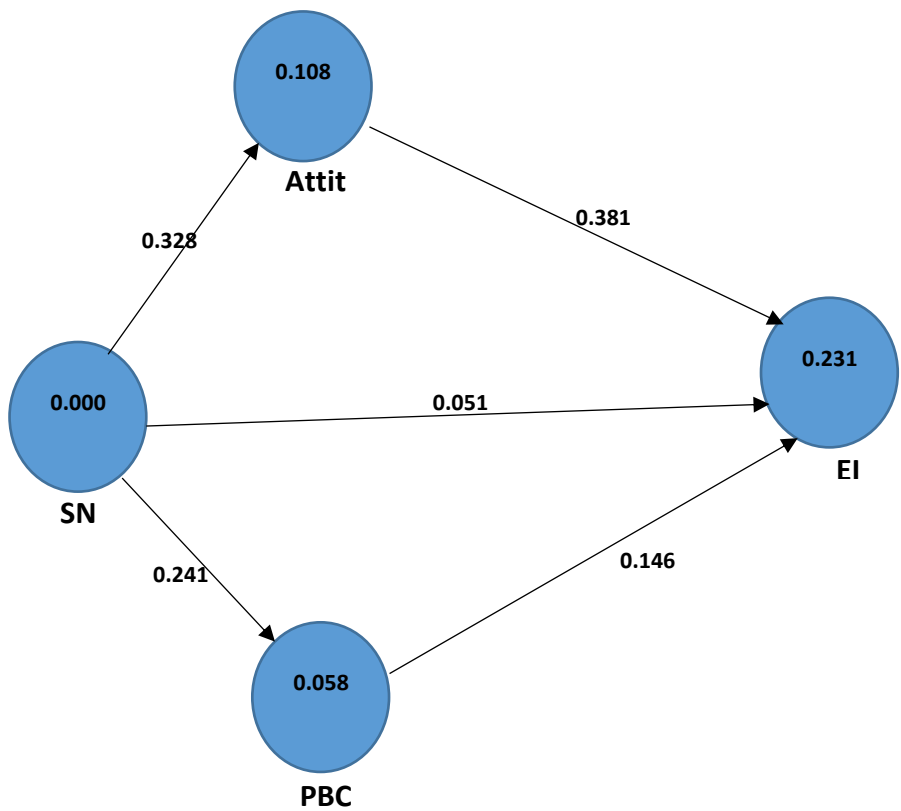

Figure 2: Structural Research Model

In order to establish the stability and significance of parameter estimates, t-values using the bootstrap procedure available in SMARTPLS were computed. Moreover, a global fit measure (the GoF index) was calculated as a geometric mean of the average communality and the average $R^2$. In this case the GoF is reported as an acceptable 0.42, which is above the cut off value of 0.36 (Wetzels et al. 2009, Tenenhaus et al. 2005). Since the objective of PLS is to maximize variance explained rather than fit, prediction-oriented measures such as $R^2$ are most appropriately used to evaluate PLS models (Chin, 1998). As shown in Figure 2 the $R^2$ for entrepreneurship intention (EI) is 0.231, suggesting that the proposed theoretical model explains 23% per cent of the variance of the construct, which is a satisfactory level of predictability (Dayan et al. 2016). The cross-validated redundancy approach to determining the $Q^2$ statistic (Stone 1974, Geisser 1975) was also examined to assess the predictive quality of the model. Results indicate a $Q^2 = 0.219$. A reported $Q2 > 0$ suggests that the model has predictive relevance (Hair et al. 2016).

Similarly, the $R^2$ for attitude was 0.108 and for perceived behavioural control (PBC) it was .058.

### TABLE 3. Path Coefficients

| | Original Sample (O) | Sample Mean (M) | Standard Deviation (STDEV) | T Statistics (\|O/STERR\|) | P value | Hypothesis |
|---|---|---|---|---|---|---|
| **Attit -> EI** | 0.381*** | 0.383 | 0.041 | 9.094 | 0.000 | **H1** |
| **PBC -> EI** | 0.146*** | 0.147 | 0.047 | 3.069 | 0.002 | **H2** |
| **SN -> EI** | 0.051 $^{ns}$ | 0.051 | 0.046 | 1.102 | 0.270 | **H3** |
| **SN -> Attit** | 0.328*** | 0.331 | 0.041 | 7.868 | 0.000 | **H4** |
| **SN -> PBC** | 0.241*** | 0.244 | 0.044 | 5.447 | 0.000 | **H5** |

*P<0.10, **P<0.05, ***P<0.01, ns is not significant

The findings generally support the conceptual model of the research. Results support the majority of hypothesized relationships. Table 3 shows the estimated standardized parameters for the causal paths. Apart from subjective norm (H3) (Standardized Estimate=0.050, P= 0.270 which is not significant), the suggested factors positively and directly affect entrepreneurship intention in the UAE, namely; attitude (H1) (Standardized

Estimate=0.380, P< 0.01) and perceived behavioural control (H2) (Standardized Estimate=0.146, P< 0.01).

Furthermore, subjective norm positively affects both attitude (H4) (Standardized Estimate=0.328, P< 0.01) and perceived behavioural control (H5) (Standardized Estimate=0.241, P< 0.01).

Since the theorized causal effects of the suggested factors may be direct or indirect i.e., mediated via the effects of other variables, or both, the total causal effects are also computed. Table 4 shows the total effects of the suggested factors.

**TABLE 4. Total Effect**

|  | Original Sample (O) | Sample Mean (M) | Standard Deviation (STDEV) | Standard Error (STERR) | T Statistics (|O/STERR|) | P Value |
|---|---|---|---|---|---|---|
| **Attit -> EI** | 0.380*** | 0.383 | 0.041 | 0.041 | 9.094 | 0.000 |
| **PBC -> EI** | 0.146*** | 0.147 | 0.047 | 0.047 | 3.069 | 0.002 |
| **SN -> EI** | 0.211*** | 0.214 | 0.047 | 0.047 | 4.459 | 0.000 |
| **SN -> Attit** | 0.328*** | 0.331 | 0.041 | 0.041 | 7.868 | 0.000 |
| **SN -> PBC** | 0.241*** | 0.244 | 0.044 | 0.044 | 5.447 | 0.000 |

*P<0.10, **P<0.05, ***P<0.01, ns is not significant

Table 4 above indicates that the total effect of SN on EI is positive and significant. Similar conclusions also hold for PBC and Attitude. The authors identify that subjective norm has marginally insignificant direct impact on entrepreneurship intention. However, subjective norm indirectly affects entrepreneurship intention through both attitude and perceived behavioural control (Total Standardized Estimate=0.211, P< 0.01).

A secondary aim of this research was to measure the effect and relations of socio-demographic variables on the research variables. The categorical demographic variables were coded as binary variables (1,0) for female, male; (1,0) for prior entrepreneurship education or not; (1,0) for previous entrepreneurship experience or not.The model displayed in Figure 3 is an exemplar for the case of gender.

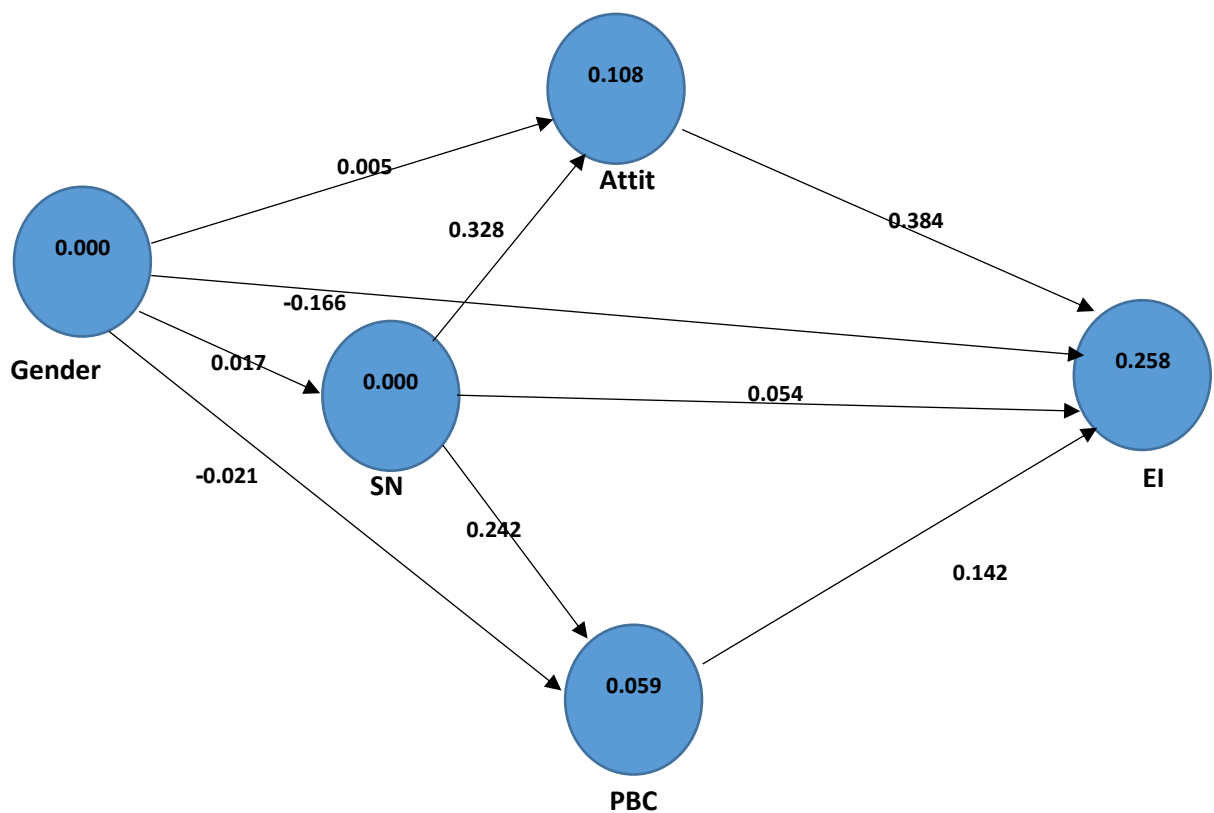

Figure 3: Structural Model- Gender

Apart from entrepreneurship intention which is negatively affected by gender (Standardized Estimate= 0.166, P< 0.01), the other suggested factors are not affected by gender, namely, attitude (Standardized Estimate=0.005, P= 0.886), perceived behavioural control (Standardized Estimate= -0.021, P= 0.613) and subjective norms (Standardized Estimate=0.017, P= 0.696).

Moreover, the results show that there is a negative total (indirect) effect of gender on entrepreneurial intention through the direct effect on attitude, subjective norm and perceived behavioral control. Accordingly, hypothesis H7a is supported, though not H7b,c & d. Indeed, the result shows that gender has a significant negative direct and indirect effect on EI. In addition to the above model for gender, similar models were developed to measure the hypothesized influence of age, entrepreneurship experience, entrepreneurship education. The results for all models displayed an $R^2$ between 0.242 and 0.258, $Q^2 > 0$ and acceptable GoF ranging from 0.41 to 0.47.

Analysis showed that entrepreneurship intention (Standardized Estimate=-0.075, P< 0.05) is positively and directly affected by age, supporting hypothesis H8a. However, attitude (Standardized Estimate= -0.004, P= 0.911) perceived behavioral control (Standardized Estimate= 0.006, P= 0.861), and subjective norms (Standardized Estimate=0.011, P= 0.779) are not influenced by age.

Similarly, a model was developed to measure the effect of prior entrepreneurship experience. The results showed that entrepreneurship intention (Standardized Estimate=-0.079, P< 0.05) and perceived behavioral control (Standardized Estimate=0.077, P= 0.057) are positively and directly affected by entrepreneurship experience, supporting hypotheses H9a & H9d. However, attitude (Standardized Estimate= 0.033, P= 0.388), and subjective norms (Standardized Estimate=0.058, P= 0.203) are not directly influenced by entrepreneurship experience. The total effects results support the view that entrepreneurship experience has both a direct and an indirect effect on entrepreneurship intention.

However, when measuring the effect of education, the study results did not support the hypothesis that education influences entrepreneurial intention, attitude and perceived behavioral control (H6a, b & c). However, results did support the hypothesized view that entrepreneurship education does not directly influence subjective norm (H6d).

## Discussion

This study aimed to investigate the intention among a sample of UAE national youth to start their own entrepreneurial ventures by examining the factors that influence the choice of entrepreneurship as a career option. More specifically, the study adopted the theory of planned behavior (TPB) based entrepreneurial intention model suggested by Liñán & Chen (2009), while also considering the unique context of the UAE.

The study findings generally support the conceptual model of the research, and hence support the applicability, robustness and validity of the TPB to investigate entrepreneurial intention among UAE national youth. The applicability of this model to entrepreneurship has received wide support in the past (Kautonen et al., 2015; Liñán & Chen, 2009; Moriano et al., 2011; Rueda et al., 2015; Turker & Selcuk, 2009) and it has also been used to explain the entrepreneurial intention of students in the USA (Autio et al., 2001), the Netherlands (Van Gelderen et al., 2008), Russia (Tkachev & Kolvereid, 1999), Spain, Taiwan (Liñán & Chen, 2009) and Saudi Arabia (Almobaireek & Manolova, 2012) among other countries and contexts. Current findings add further support to the view that the TPB may be useful as a *"culture-universal theory",* to predict career intentions in any country (Moriano et al., 2011, p. 2).

The general results of the present study support the direct and positive effects of attitude and perceived behavioral control on the entrepreneurial intention of UAE national youth, whereas subjective norm appears to have an indirect effect on entrepreneurial intention through its direct effect on attitude and perceived behavioral control. This is aligned with the findings by Liñán & Chen, (2009). Although the study results support the past research findings that the subjective norm has an insignificant and weak direct relationship with entrepreneurial intention (Armitage & Conner, 2001; Kautonen et al., 2015; Matlay et al., 2012; Moriano et al., 2011; Schlaegel & Koenig, 2014), it does not necessarily support the claim that it is the least important predictor of students' entrepreneurial intention, as suggested by Moriano et al. (2011). As noted above, subjective norm in this study, is found to have a relatively high effect on both attitude (standard regression weights = .328) and perceived behavioral control (standard regression weights = .241), which in turn affects entrepreneurial intention.

Accordingly, this variable appears to be a relatively strong indirect predictor of entrepreneurial intention among this sample of UAE national youth.

In line with past research findings attitude appears to have a positive and direct effect on the entrepreneurial intention of UAE national youth (Autio et al., 2001; Liñán & Chen, 2009; Matlay et al., 2012; Moriano et al., 2011; Rueda et al., 2015; Van Gelderen et al., 2008). Indeed, this construct is suggested by previous research to be the strongest predictor of entrepreneurial intention (Ajzen, 1991; Moriano et al., 2011); a view supported by the current results.

Many researchers emphasize the role of perceived behavioral control on predicting entrepreneurial intention, especially for student samples (Armitage & Conner, 2001; Kautonen et al., 2015; Matlay et al., 2012; Schlaegel & Koenig, 2014). Although it is the second most significant predictor of entrepreneurial intention, after attitude, current results identify perceived behavioral control as a moderate to weak predictor of entrepreneurial intention for this UAE sample.

In relation to controlled variable effects, the general study supports Liñán & Chen's claim (2009) that these variables may have significant but weak influence on youth entrepreneurial intention (Liñán & Chen, 2009). Indeed, the indirect effects of these variables are also found by many past researchers (Almobaireek & Manolova, 2012; Yang, 2013; Kolvereid & Isaksen, 2006). However, the study also found some degree of a significant but weak direct effect on entrepreneurial intention of all these controlled variables except entrepreneurship education. Education appears to have no significant effects on entrepreneurial intention or its antecedents in the UAE context. Thus, it not only has no significant effect on subjective norms, as this study proposes, but also has no effect on any of the theory's other constructs.

This study employed the "Entrepreneurial Intention Questionnaire" developed by Liñán and Chen (2009). As supported by Liñán and Chen (2009) and Rueda et al. (2015), the EIQ appears to be a valid instrument with which to study entrepreneurship intention in the UAE context. Results further support the use of measurement instrument as a valid, reliable and standardized measurement in the study of entrepreneurial intention, whatever the context.

## Practical Implications

On a practical note, the study has many implications and recommendations for both educators and policy makers seeking to enhance entrepreneurial intention among UAE national youth.

Firstly, as attitude appears to be the strongest predictor of entrepreneurship intention in this group, more focus should be given to enhancing UAE national youth entrepreneurial intention through improving attitudes to entrepreneurship. Education is important in this regard. However, experience of entrepreneurship education in this study appears not to have any significant effect on national youth attitude to entrepreneurship as career choice. This finding should prompt government universities to fully evaluate the nature and quality of entrepreneurship education in the higher education institutions of the UAE and determine how it can be improved. Second, because the subjective norm in this study appears to strongly affect both attitude and perceived behavioral control, there is great need to include UAE national youth families and wider communities in any strategies or initiatives aimed to enhance youth EI. Additionally, to enhance the role of perceived behavioral control, thus enhancing national youth confidence as potential entrepreneurs, the study suggests that UAE national youth entrepreneurship experience could be broadened, perhaps through expanding internships or attaching them to successful start-up enterprises locally or abroad.

## Limitations and Future Research

The research findings reveal certain specific limitations that should be highlighted. First, in this study, UAE national youth entrepreneurship intention was examined based on subgroup of UAE national undergraduate's students. Indeed, for practical reasons, and with support from the literature, the study limits its population to business and engineering students (Pratheeba, 2014; Souitaris et al., 2007). Accordingly, this limits the strength and generalizability of results, as with any study which examines a specific target population. Also the current study does not take into account the influence that institutional barriers can have on entrepreneurship (Yukhanaev, Fallon, Baranchenko & Anisimova, 2015). Rather it only examines the likely consequence of such barriers on the attitude, perceived behavioral control subjective norm of participants.

This study highlighted many interesting avenues for future research. First, the study found that the control variables of age, gender, and entrepreneurship experience directly affect UAE national youth entrepreneurial intention, which is contrary to claims that these variables should only affect entrepreneurship intention indirectly through the theory's main three constructs (Ajzen, 1991). Accordingly, future researchers should further investigate this issue. Second, this study targets only government universities in its investigations. Future research is encouraged to target also private and other universities to strengthen the generalizability of the research findings. Moreover, targeting other universities and colleges to investigate the EI of senior students in different college majors is recommended. Third, this quantitative study used a survey questionnaire to collect and investigate the study data. Future researchers are encouraged to apply a mix of quantitative and qualitative methodologies to strengthen the research findings and offer further insight regarding the EI of youth and the factors affecting it.